\documentclass[%
    reprint,
superscriptaddress,
showpacs,
 amsmath,amssymb,
prb,
]{revtex4-1}

\usepackage[colorlinks,citecolor=black,linkcolor=black,urlcolor=blue,bookmarks=true,hypertexnames=true]{hyperref} 
\usepackage{graphicx}% Include figure files
\usepackage{color}
\usepackage{dcolumn}% Align table columns on decimal point
\usepackage{tabularx}

\usepackage[T1]{fontenc}
\usepackage{braket}

\begin{document}

\title{Temperature-dependence of $T_1$ spin relaxation of single NV centers in nanodiamonds}% Force line breaks with \\

\author{T. de Guillebon}
\author{B. Vindolet}
\author{J.-F. Roch}
\affiliation{Universit\'e Paris-Saclay, CNRS, ENS Paris-Saclay, CentraleSup\'elec, LuMIn, 91405, Orsay, France}
\author{V. Jacques}
\affiliation{Laboratoire Charles Coulomb, Universit\'e de Montpellier and CNRS, 34095 Montpellier, France}

\author{L. Rondin}
\email{loic.rondin@universite-paris-saclay.fr}
\affiliation{Universit\'e Paris-Saclay, CNRS, ENS Paris-Saclay, CentraleSup\'elec, LuMIn, 91405, Orsay, France}

\date{\today}

\newcommand{\cor}[1]{{\color{red}{#1}}}

\begin{abstract}
We report the experimental study of the temperature-dependence of the longitudinal spin relaxation time $T_1$ of single Nitrogen-Vacancy (NV) centers hosted in nanodiamonds. To determine the relaxation mechanisms at stake, measurements of the $T_1$ relaxation time are performed for a set of individual NV centers both at room and cryogenic temperatures. The results are consistant with a temperature-dependent relaxation process which is attributed to a thermally-activated magnetic noise produced by paramagnetic impurities lying on the nanodiamond surface. These results confirm the existence of surface-induced spin relaxation processes occurring in nanodiamonds, which are relevant for future developments of sensitive nanoscale NV-based quantum sensors.
\end{abstract}

\maketitle

The development of the Nitrogen-Vacancy (NV) center in diamond as a quantum sensor has triggered numerous applications, from imaging magnetism at the nanoscale~\cite{RONDIN2014} to nuclear magnetic resonance on small spin ensembles and proteins~\cite{Lovchinsky2016,Lovchinsky2017,Abobeih2019}.
In that context, spin relaxation times of these quantum sensors play a central role. Besides setting an intrinsic limit to the sensor sensitivity, variations of the spin relaxation properties can also be exploited for noise sensing at the nanoscale~\cite{DEGEN2017}. For instance, measurements of the longitudinal spin relaxation time $T_1$ can be used for studying conduction in metals~\cite{KOLKOWITZ2015,ARIYARATNE2018}, noise spectroscopy in 2D systems~\cite{AGARWAL2017} and magnetic resonance~\cite{vanderSar2015,WOOD2016,SCHAFER2014,SADZAK2018}.

Identifying and understanding the parameters impacting the $T_1$ relaxation time is therefore essential for NV-based sensing technologies. For NV centers isolated in bulk diamond, phononic induced spin-lattice relaxation has been highlighted as the dominant process that limits the value of $T_1$ at room temperature~\cite{JARMOLA2012,ASTNER2018,NORAMBUENA2018}. At the same time, cross-relaxation due to the coupling with ancillary spins species~\cite{JARMOLA2012,BAUCH2019}, as well as the influence of electric and magnetic field noises~\cite{JARMOLA2012,TETIENNE2013,MYERS2018}, also need to be taken into account. 
Conversely, NV centers close to a diamond surface, \textit{i.e.} both in nanodiamonds and shallow in bulk, have $T_1$-times that are orders of magnitude shorter than those measured in bulk, thus introducing a stringent limit to their application for quantum sensing. Hence, numerous recent studies have addressed the physical mechanisms at the origin of the short $T_1$-time observed for NV centers under these conditions~\cite{ROSSKOPF2014,MYERS2014,Romach2015,MYERS2018,KIM2015,Sangtawesin2019}.

In the case of NV centers in nanodiamonds, it has been shown that the $T_1$-time strongly depends on the size of the host nanodiamond~\cite{TETIENNE2013}.  These results highlighted the impact of surface-related noise on the relaxation time, which has been attributed to magnetic fluctuations induced by dipolar coupling between paramagnetic impurities lying on the nanodiamond surface. Since such dipolar interaction is expected to be temperature independent, experiments performed at cryogenic temperature should not yield an improvement of the $T_1$-time of NV centers in nanodiamonds, conversely to what is observed for NV centers in a bulk diamond~\cite{JARMOLA2012,ASTNER2018,NORAMBUENA2018}. To verify this hypothesis, we investigate the effect of temperature on the $T_1$ relaxation time of single NV centers hosted in nanodiamonds. More precisely, we focus on the impact of magnetic field noise by addressing specifically the relaxation between the $m_s=0$ state and $m_s=\pm1$ states.
This work provides new insights on surface-induced spin relaxation processes occurring in nanodiamonds, which are relevant for future developments of sensitive nanoscale NV-based quantum sensors~\cite{McGuinness2013,TETIENNE2015,Rondin2010}.

%%%%%%%%%%%%%%%%%%%%%%%%%%%%%%%%%%%%%%%%%%%%%%%%%%%%%%%%%%%%%%%%%%%%%%%%%%%%%%%%%%%%%%%%%%%
%%%                                                                                     %%%
%%%                                 FIGURE 1                                            %%%
%%%                                                                                     %%%
%%%%%%%%%%%%%%%%%%%%%%%%%%%%%%%%%%%%%%%%%%%%%%%%%%%%%%%%%%%%%%%%%%%%%%%%%%%%%%%%%%%%%%%%%%%
\begin{figure}[ht!]
%\vspace*{-2.2cm}
\includegraphics[width=8.6cm]{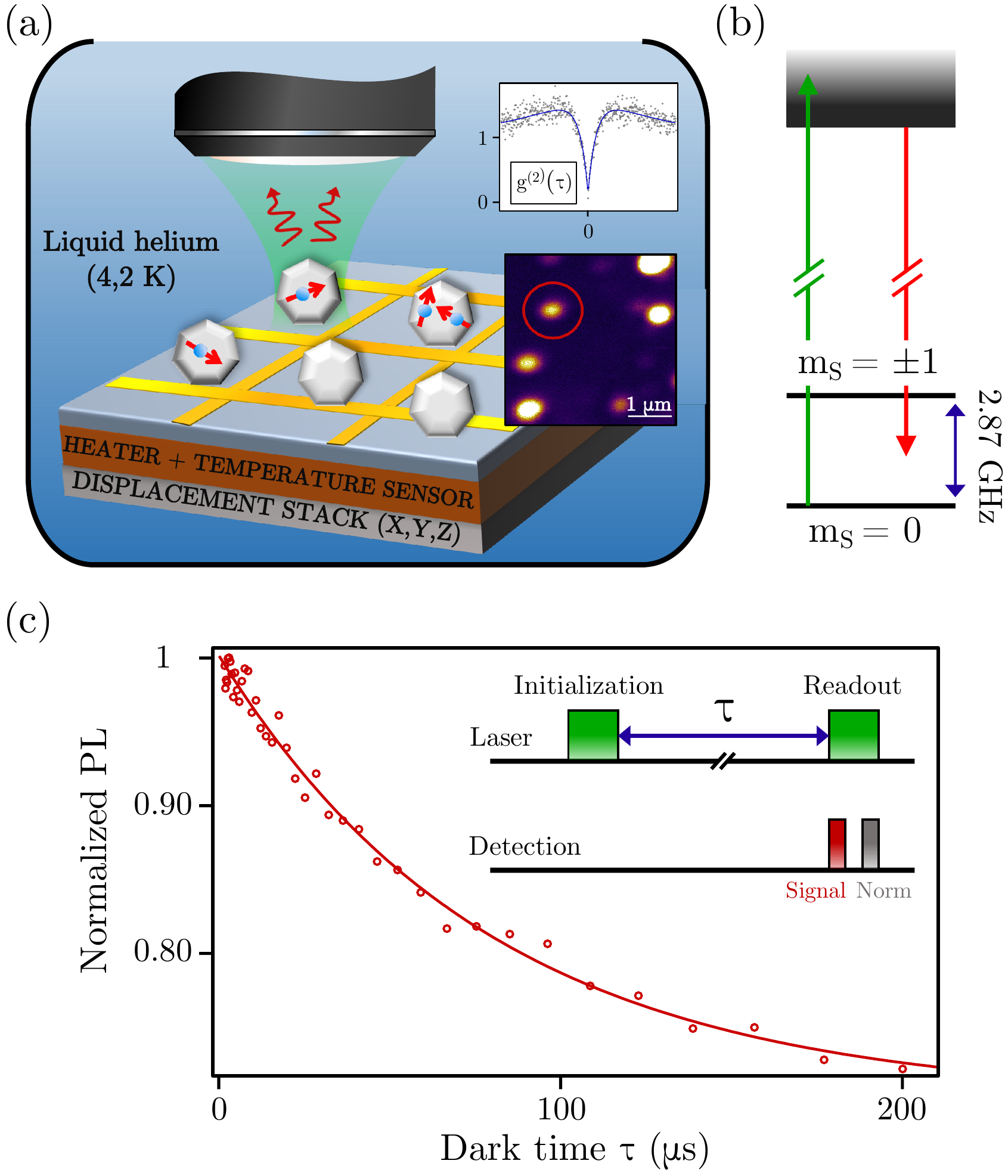}
%\vspace*{-3.9cm}
\caption{(a) Experimental apparatus. A scanning confocal microscope is used to identify single NV centers in nanodiamonds which are deposited on a quartz coverslip patterned with a golden grid. A typical photoluminescence raster scan of the sample is shown in the lower inset, where the red circle indicate a single NV center. The unicity of the emitter is verified by measuring the second-order correlation function $g^{(2)}(\tau)$. As shown in the upper inset, $g^{(2)}(0)<0.5$ is the signature of a single quantum emitter. (b) Simplified energy-level structure of the NV defect. (c) Typical $T_1$-time measurement at room temperature. The solid line is an exponential fit, which yields $T_{1}^{RT}=82.6\pm6.6$~$\mu$s. The inset presents the experimental pulse sequence.}
\label{fig1}
\end{figure}

To investigate the $T_1$-time of single NV centers, a solution of electron-irradiated type-Ib diamond nanocrystals~\cite{Rondin2010} (50-nm average size) was spin-coated on a quartz coverslip. To ensure the reproducibility of the experiment, a golden grid was patterned on the coverslip prior to the deposition of the nanodiamonds, allowing the identification of each nanodiamond location [Fig.~\ref{fig1}(a)].
The NV center optical properties are addressed using a customized confocal microscope. A Hanbury Brown and Twiss interferometer is used to record the statistics of photon correlations and unambiguously identify nanodiamonds hosting a single NV defect (upper inset of Fig.~\ref{fig1}(a)). Only such nanodiamonds are used in the present study to avoid ensemble averaging effects~\cite{TETIENNE2013}.

%%%%%%%%%%%%%%%%%%%%%%%%%%%%%%%%%%%%%%%%%%%%%%%%%%%%%%%%%%%%%%%%%%%%%%%%%%%%%%%%%%%%%%%%%%%
%%%                                                                                    %%%
%%%                                 FIGURE 2                                            %%%
%%%                                                                                     %%%
%%%%%%%%%%%%%%%%%%%%%%%%%%%%%%%%%%%%%%%%%%%%%%%%%%%%%%%%%%%%%%%%%%%%%%%%%%%%%%%%%%%%%%%%%%%
\begin{figure}
%\hspace{-0.3cm}
\centering
\includegraphics[width=8.6cm]{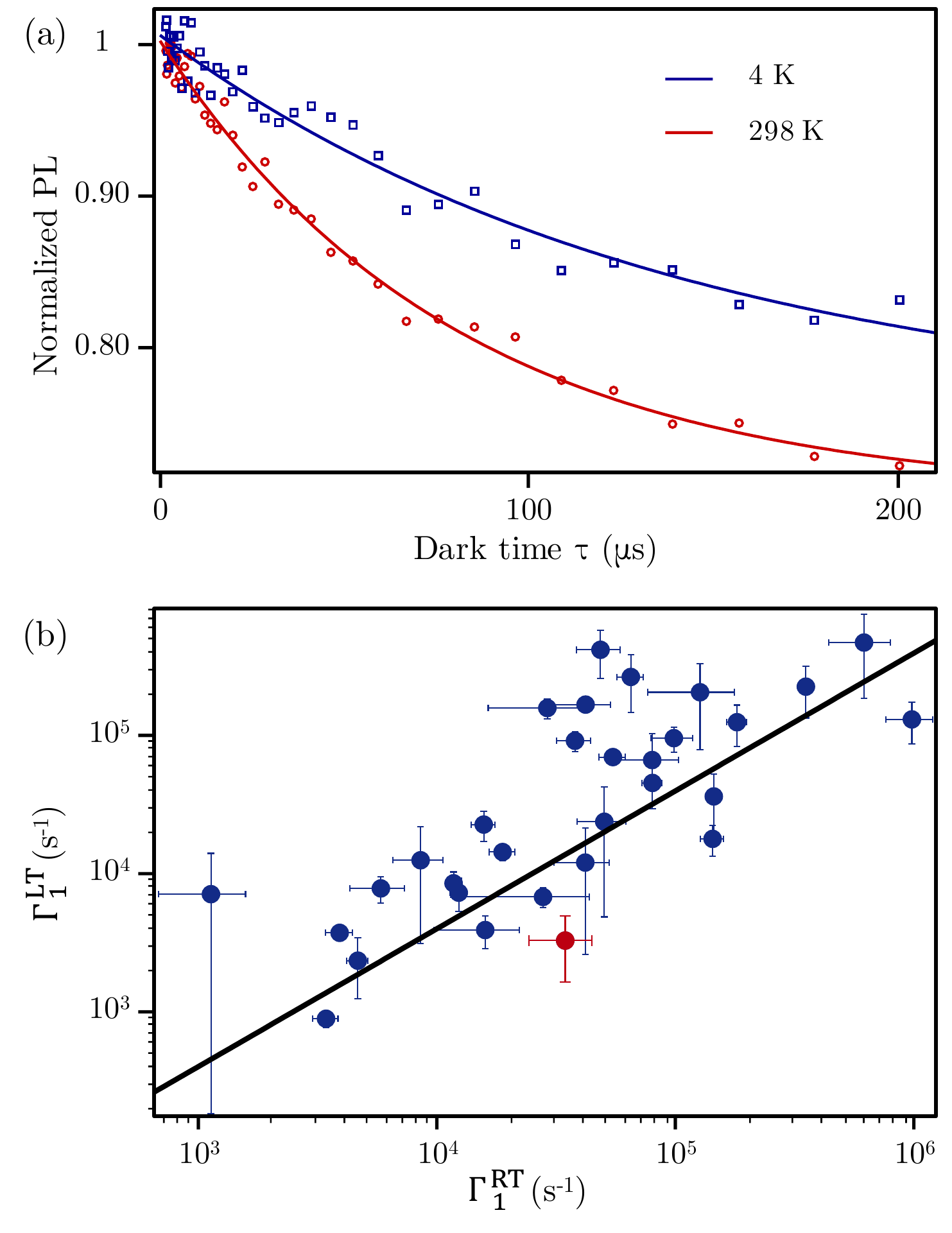}
\caption{(a) Spin relaxation rate measurements for the same single NV center at room (298~K, red) and cryogenic temperature (4~K, blue). The solid lines are exponential fits that yield $T_{1}^{RT}=82.6\pm6.6$~$\mu$s and $T_{1}^{LT}=137\pm37$~$\mu$s. (b) Spin relaxation rate $\Gamma_1=1/T_1$ at cryogenic temperature (4~K) as a function of the measured relaxation rate at room temperature (298 K) for the same point defect, measurement done on a set of thirty-one NV centers in distinct nanodiamonds. The correlation between the two parameters is plotted in a log-log scale. The error bars are extracted from the relaxation rate fit uncertainties. The red point corresponds to the NV center which behavior is reported in Fig.~\ref{G_vs_T}. A linear fit (solid black line) gives the ratio $\Gamma_1^\text{LT}/\Gamma_1^\text{RT}=0.41\pm0.1$, from which we extract an average activation temperature of $3.8\pm 1.1$~K, following the model developed in the main text.}
\label{GLT_vs_GRT}
\end{figure}

For an identified single NV center in a given nanodiamond, we measure the relaxation rate $\Gamma_1=1/T_1$ between the $m_s=0$ and $m_s=\pm1$ spin sublevels [Fig.~\ref{fig1}(b)]. As a consequence, we address specifically the relaxation of single quantum transitions, \textit{i.e.} between states of $\Delta m_s=\pm1$ which is primarily due to magnetic field noise~\cite{MYERS2018}.
Using a continuous wave 532-nm laser excitation, then chopped in pulses using an acousto-optic modulator, the measurement is performed with the pulse sequence shown in the inset of Fig.~\ref{fig1}(c). The NV center is initially polarized into the $m_s=0$ state by an optical pulse of 3~$\mu$s duration. Using a second optical pulse, the spin state is then read out, after a variable dark time of duration $\tau$, causing the system to decay into a mixture of $m_s = 0, +1$ and $-1$ spin sublevels. The collected photoluminescence signal as a function of the parameter $\tau$ exhibits an exponential decay with a characteristic time corresponding to the relaxation time $T_1$~\cite{TETIENNE2013}. Note that all the experiments presented in the following are done without any bias magnetic field. A typical experiment done at 298~K for a single NV defect hosted in a nanodiamond of 10 nm size is shown in Fig.~\ref{fig1}(c) (red circles). An exponential fit leads to the value $T_{1}^\text{RT}=82.7\pm6.6$~$\mu$s, at room temperature. This result is in agreement with expected values for NV centers in nanodiamonds~\cite{TETIENNE2013,McGuinness2013}, and is orders of magnitude smaller than the typical value measured for NV centers in bulk diamond. 
To record the temperature dependence of the $T_1$-time, the experimental apparatus is inserted in a cryostat (Attocube) designed to reach cryogenic temperature, using helium as an exchange gas at a pressure of about 50 mbar.
The patterned grid is used to measure the $T_1$-time on the same NV center previously studied at room temperature. Measurements performed for the same NV defect at 298~K and 4~K are shown in Fig.~\ref{GLT_vs_GRT}(a). At low temperature, we observe that the $T_1$ relaxation time increases to $137\pm37$~$\mu$s, significally above the error bars. The difference between the measurements performed at room and liquid-helium temperatures thus reveals the existence of thermally activated mechanisms on the NV spin relaxation rate. 

To confirm this effect, we then measure the variation of the $T_1$-time with temperature for a set of about thirty individual NV centers.
Figure~\ref{GLT_vs_GRT}(b) shows the relaxation rate $\Gamma_1^{LT}$ of the NV centers at low temperature as a function of the relaxation rate $\Gamma_1^{RT}$ at room temperature. 
First, we note that the room-temperature relaxation rate  $\Gamma_1^{RT}$ spreads over more than two orders of magnitude. This is due to the size distribution of the nanodiamonds, which are measured from $\approx$~10 to 70~nm using an atomic force microscope.
Besides, these size measurements for the whole set of studied nanodiamonds allow us to confirm the findings of reference~[\onlinecite{TETIENNE2013}] (data not shown): the relaxation rate $\Gamma_1$ is correlated to the nanodiamond size, the shortest rates being observed for the larger nanodiamonds. This effect can simply be explained by the fact that the nanodiamond size impacts the distance of the NV center to the surface, and, consequently, the strength of surface-related magnetic noise.
Second, while decreasing the temperature, we observe a decrease of $\Gamma_1$ for most of the nanodiamonds, regardless of their size. To get a trend of this effect, we use a linear fit (solid black line in figure~\ref{GLT_vs_GRT}(b)) and estimate the average ratio $\Gamma_1^\text{LT}/\Gamma_1^\text{RT}=0.41\pm0.1$. Since this ratio is smaller than unity, these measurements give evidence for the existence of temperature-dependent relaxation processes that cannot be neglected.
Finally, we also observe that the increase of the $T_1$-time seen by cooling the nanodiamonds is very moderate compared to the one observed in bulk diamond, where relaxation times increase from ms to tens of seconds by reaching cryogenic temperature~\cite{NORAMBUENA2018}. Similar increases have also been observed for shallow NV centers located at a distance smaller than 10 nm from the crystal surface~\cite{ROSSKOPF2014,Romach2015}. These values are far greater than those measured for NV centers in nanodiamonds, such as the ones shown in figure~\ref{GLT_vs_GRT}(b).

To model the temperature dependence of the $T_1$-time, we assume that the decay between the population of the spin sublevels $m_s=0$ and $m_s=\pm1$ is still induced by magnetic noise generated by paramagnetic impurities on the surface of the nanodiamond. We model this noise as a random magnetic field, which is created by an ensemble of fluctuating spins with a Lorentzian spectral density. We then introduce the correlation time $\tau_c$, which corresponds, in the case of magnetic noise generated by a spin bath, to the typical reversal time of the spins~\cite{TETIENNE2013,ROSSKOPF2014,Romach2015}. The relaxation rate of the NV center, at its electron spin resonance frequency $\omega_0/(2\pi)\approx 2.87$~GHz, then reads~\cite{TETIENNE2013,ROSSKOPF2014,Romach2015}

\begin{equation}
\Gamma_{1}=\frac{1}{T_1}=\Delta \frac{\tau_c^{-1}}{\tau_c^{-2}+\omega_0^2}\, ,
\end{equation}
where $\Delta$ represents the strength of the interaction between the spin bath and the NV center. This parameter is proportional to the magnetic noise at the NV center location, which drastically depends on the NV position inside the nanodiamond~\cite{TETIENNE2013}. 
We then consider two contributions in the parameter $\tau_c$: a dipolar component, associated with a fluctuation rate $R_\text{dip}$, due to intra-bath coupling between fluctuating magnetic dipoles, and a temperature-dependent component, associated with a fluctuation rate $R_T$. 
We can thus write $\tau_c^{-1}=R_T+R_\text{dip}$, where the dipolar component $R_\text{dip}$ is temperature-independent.

\begin{figure}[ht!]
%\hspace{-0.3cm}
\includegraphics[width=8.6cm]{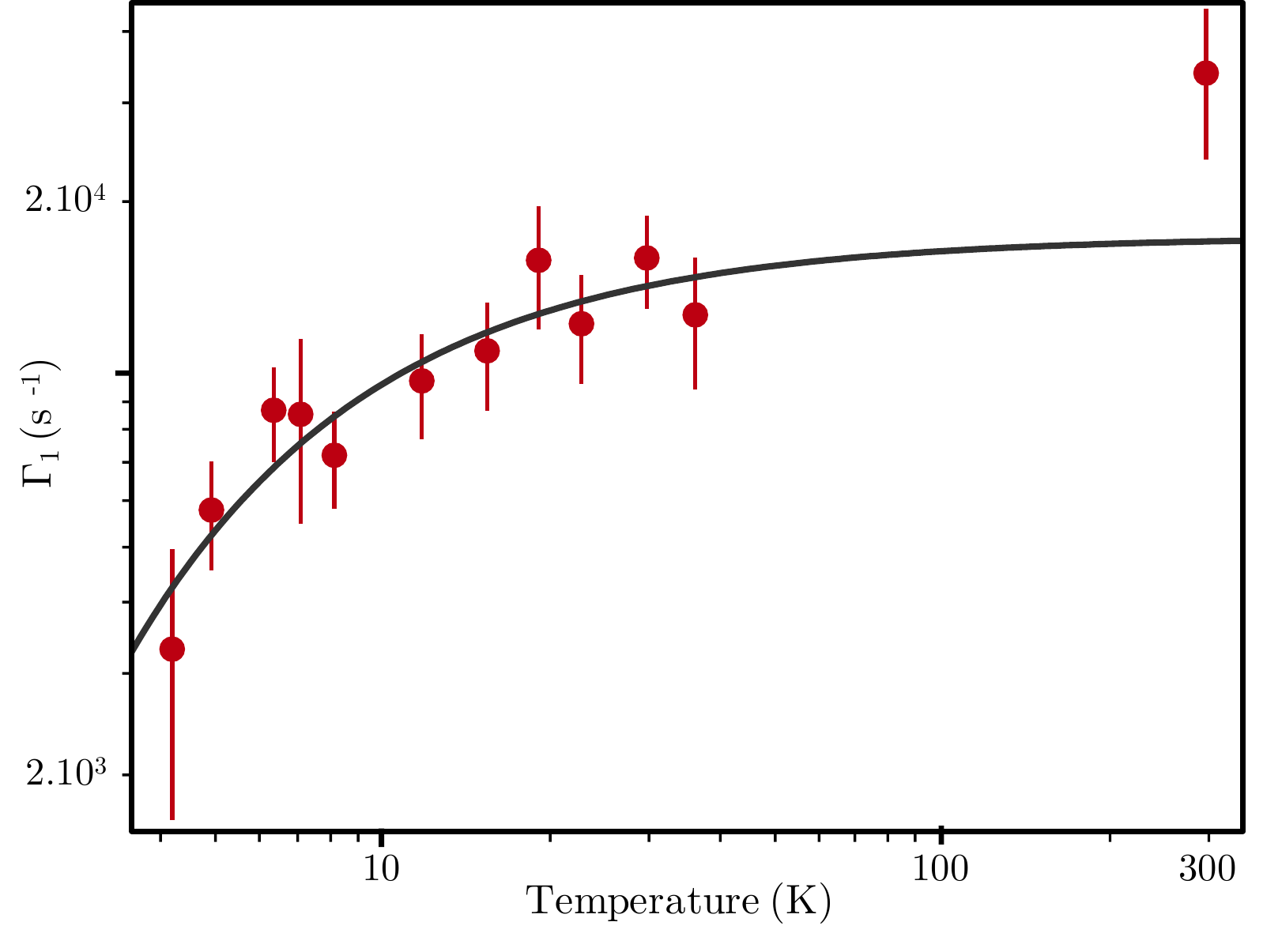}
\caption{Relaxation rate of a single NV center in a 20-nm nanodiamond as a function of the temperature, plotted in log-log scale. The solid black line represents a fit following equation~(\ref{eq:G1bis}), the data being weighted by their uncertainties. We extract an activation temperature of $T_{a}=6.0 \pm 1.2$~K and a high temperature relaxation rate of $\Gamma_\infty=(17\pm2)\cdot10^3$~s$^{-1}$.} %In addition, the fit provides that the high temperature rate $R_\infty$ is small compare to $\omega_0$.}
\label{G_vs_T}
\end{figure}

To investigate the relative contribution of these two components, we then studied the $T_1$ relaxation time of a typical NV center in a nanodiamond as a function of the temperature $T$.
To this end, we used a resistive heater placed below the sample holder [Fig.~\ref{fig1}(a)] to control the sample temperature between 4~K and $\sim 50$~K. Figure~\ref{G_vs_T} shows the measured relaxation rate $\Gamma_1$ as a function of the temperature. 
We do not observe a saturation of the $\Gamma_1$ relaxation rate at the lowest temperature that can be reached using our cryogenic setup. This behavior indicates that, even at 4~K, relaxation is still dominated by a temperature-dependent process. We thus conclude that the magnetic noise component induced by the intra-bath coupling between fluctuating magnetic dipoles on the diamond surface is negligible for the whole range of investigated temperatures.
As a result, we rewrite the correlation time as $\tau_c^{-1}\approx R_T$. In addition, given the saturation of the relaxation rate at high temperature (Fig.~\ref{G_vs_T}), we postulate a simple model of a thermally activated fluctuation rate corresponding to $R_T (T)=R_\infty \exp\left(-{T_a}/{T} \right)$, with $T_a$ the activation temperature of the process, and $R_\infty$ the fluctuation rate at high temperature. 
This model thus leads to
\begin{equation}
    \Gamma_{1}=\frac{1}{T_1}\approx\Delta \frac{R_\infty e^{-T_a/T}}{\left(  R_\infty e^{-T_a/T}\right)^2+\omega_0^2}\, .
    \label{eq:G1}
\end{equation}

This equation is then used to fit the data shown in figure~\ref{G_vs_T}. By evaluating the convergence of the fit, we observe that $R_\infty$ and $\Delta$ are strongly correlated and cannot be determined independently. However, $R_\infty$ is required to be at least one order of magnitude smaller than $\omega_0$. As a result, we can assume, for the whole range of temperatures studied, that the relaxation rate can be written as 

\begin{equation}
    \Gamma_{1}\approx\Gamma_\infty  e^{-T_a/T}\, ,
    \label{eq:G1bis}
\end{equation}

where $\Gamma_\infty=(\Delta R_\infty)/\omega_0^2$ represents the relaxation rate at high temperature. Finally, weighting the data by their uncertainties, the fit by the equation (\ref{eq:G1bis})  gives an estimate of the activation temperature for this given NV center to be $T_a=6.0\pm1.2$~K. 

%%%%%%%%%%%%%%%%%%%%%%%%%%%%%%%%%%%%%%%%%%%%%%%%%%%%%%%%%%%%%%%%%%%%%%%%%%%%%%%%%%%%%%%%%%%
%%%                                                                                     %%%
%%%                                 FIGURE 3                                            %%%
%%%                                                                                     %%%
%%%%%%%%%%%%%%%%%%%%%%%%%%%%%%%%%%%%%%%%%%%%%%%%%%%%%%%%%%%%%%%%%%%%%%%%%%%%%%%%%%%%%%%%%%%

Based on this result, we discuss the pertinence of this model for the whole set of single NV centers. To do so, we assume that all nanodiamonds have identical surface states. Thus, the observed spread in the $\Gamma_1$ relaxation rate which is observed in figure~\ref{GLT_vs_GRT}(b), is ascribed to a change in the temperature-independent interaction strength $\Delta$, determined by the NV-to-surface distance. Also, we consider the activation temperature $T_a$ to be the same for all the nanodiamonds. Finally, given the equation (\ref{eq:G1bis}), we can rewrite the ratio between the rates at low and room temperatures as

\begin{equation}
    \Gamma_1^\text{RT}/{\Gamma_1^\text{LT}}=\exp\left(\frac{T_a}{T^\text{RT}}-\frac{T_a}{T^\text{LT}}\right) \, . 
\end{equation}

From the data and the linear fit shown in figure~\ref{GLT_vs_GRT}(b), we thus determine an activation temperature of $3.8\pm 1.1$~K. This result, averaged over a set of 31 single NV centers, is in good agreement with the estimated activation temperature determined in figure~\ref{G_vs_T} from the temperature-dependent $\Gamma_1$ rate measured for a single NV center. Beyond validating the model that we introduced to explain the relaxation of NV centers in nanodiamonds, these results provide a typical energy scale of surface-related magnetic fluctuations in nanodiamonds.

Although the model cannot provide a conclusive proof of the mechanisms at stake for the NV spin relaxation, their dependence on the nanodiamond size and temperature allows drawing a few hypotheses. 
Even at the lowest temperature studied, the measured $T_1$-time in nanodiamonds remains still orders of magnitude smaller than the one measured in bulk diamond. Therefore, a direct phonon process appears unlikely to be the dominant relaxation mechanism.
More likely, the paramagnetic impurities on the diamond surface impact the relaxation of the NV defect through the generation of fluctuating magnetic fields. 
As previously discussed, a pure dipolar coupling of these paramagnetic impurities is expected to be temperature-independent contrary to the experimental behavior. Alternatively, spin fluctuations of the surface paramagnetic impurities due to their internal vibrational degrees of freedom can account for generating random magnetic fields, this mechanism being temperature-dependent. These surface defects may also play the role of charge traps and lead to magnetic noise~\cite{Sangtawesin2019} but we would then expect an activation temperature far above our observations~\cite{deSousa2007}.

In conclusion, we measured the $T_1$-time of single NV centers in nanodiamonds at both cryogenic and room temperatures. We demonstrated that this relaxation time has a non-negligible temperature-dependent component, although weak compared to the measurements previously done for NV centers in bulk diamond. We show that a simple activation model describes the temperature behavior with an activation temperature of a few kelvins. We attribute this effect to the vibrational relaxation component of a bath of paramagnetic impurities lying on the nanodiamond surface. 
Decreasing further the temperature~\cite{NORAMBUENA2018} could unveil other relaxation sources, and provide clues on the actual weight of dipolar relaxation processes, by finally accessing a temperature regime where the relaxation rate becomes temperature independent. 
Besides, it would be interesting to extend the present work to study of the effect of nanodiamond surface termination~\cite{RYAN2018} and of electric field noise~\cite{Gardill2020} on the temperature dependence of the longitudinal relaxation of NV centers. Similarly, in nanodiamonds produced by chemical vapor deposition, which present an improved surface quality~\cite{Tallaire2019}, the study of the NV center relaxation mechanisms could provide essential insights.  
Bringing a new understanding of the spin relaxation processes at stake for NV centers in nanodiamonds, our work opens the way to a better understanding of the origin of diamond surface impurities. Ultimately, the influence of these impurities on the relaxation could be reduced using surface cleaning techniques~\cite{Tsukahara2019} or could be compensated using quantum control protocols~\cite{Bluvstein2019}.

\vspace{.5 cm}
The authors acknowledge T. Hingant and L. Martinez for experimental help at the start of the project. This research is supported by the Paris \^Ile-de-France Region in the framework of DIM SIRTEQ and by C'Nano \^Ile-de-France (Nanomag contract). B. Vindolet acknowledges funding by a research grant of D\'el\'egation G\'en\'erale de l'Armement.

\bibliography{Bib}

\end{document}